\documentclass[aps,reprint,prc,showpacs,amsmath,amssymb,nofootinbib]{revtex4-1}
\usepackage{graphicx}
\usepackage{color}
\usepackage{times}
\usepackage{inputenc}
\usepackage{bm}
\usepackage{multirow}
\usepackage{float}
\usepackage{url}
\usepackage{natbib}
\usepackage{tensor}
\usepackage[colorlinks=true,citecolor=blue,urlcolor=blue,linkcolor=red]{hyperref}
\begin{document}
%%% 
\title{Constraining nuclear matter parameters and neutron star observables using PREX-2 and NICER data}
\author{S. K. Biswal$^{1}$}
\email{subratphy@gmail.com}
\author{H. C. Das $^{2,3}$}
%\email{harish.d@iopb.res.in}
\author{Ankit Kumar$^{2,3}$}
%\email{ankit.k@iopb.res.in}
\author{S. K. Patra$^{2,3}$}
%\email{patra@iopb.res.in}
\affiliation{\it $^{1}$Department of Engineering Physics, DRIEMS Autonomous Engineering College, 754022, India}
\affiliation{\it $^{2}$Institute of Physics, Sachivalaya Marg, Bhubaneswar 751005, India}
\affiliation{\it $^{3}$Homi Bhabha National Institute, Training School Complex,
Anushakti Nagar, Mumbai 400094, India}
\date{\today}
%%%%%%%%%%%%%%
\begin{abstract}
We try to constraints some of the nuclear matter parameters such as symmetry energy ($J$) and its slope ($L$) from the recent inferred data of the PREX-2. Other nuclear matter parameters are adopted from {\bf [Phys. Rev. C 85 035201 (2012), Phys. Rev. C 90 055203 (2014)]} papers and the linear correlation among them are checked  by using the Pearson's formula. We find the correlation between $J-L$, $K_\tau-J$ and $K_\tau-L$ with coefficients 0.85, 0.81 and 0.76 respectively. The neutron star properties such as mass and radius are calculated with 50 unified equation of states. The results are consistent with recently observed pulsars and NICER data except few exceptions. From the radii constraints, we find that the new NICER data allows a narrow radius range contrary to a large range of PREX-2 and the old NICER data leaving us an inconclusive determination of the neutron star radius.
\end{abstract}
%%%%%%%%%%%%%%%%%
\maketitle
%\pagenumbering{roman}
%\setcounter{page}{3}
%%%%%%%%%%%%%%%%
\section{Introduction}
\label{intro}
The neutron star (NS),  a highly dense and asymmetric nuclear system having a central density 5--6 times the nuclear saturation density \cite{Lattimer_2004}. It has a unique internal structure, where all the four fundamental forces play an essential role. Study of the NS reveals that the internal structure is more complicated because new degrees of freedom like hyperons \cite{Ambartsumyan_1960,NKGh_1985,Schaffner_1996,Biswal_2016,Fortin_2017,Bhuyan_2017,Biswalaip_2019,Biswal_2019} and quarks  are in the core \cite{Collins_1975,Orsaria_2014,Mellinger_2017}. To explore its properties, such as mass, radius and tidal deformability, etc., one has to consider the interaction between nucleons  in the form of interaction Lagrangian. This provides the equation of state (EOS), the main ingredient for the calculation of  the NS properties. 

Different formalism have been developed to calculate the EOSs of the NS. The relativistic mean-field (RMF) \cite{Miller_1972,Serot_1986,Furn_1987,Reinhard_1988,Frun_1997,Kumar_2017,Kumar_2018}, Skyrme-Hartree-Fock (SHF) \citep{Skyrme_1956,Skyrme_1958,Vautherian_1972,Chabanta_1998,Brown_1998,Stone_2007,Dutra_2012, Gogny_1980}, density-dependent RMF (DD-RMF) \cite{Typel_2005}, and point couplings \cite{Dutra_2014} formalism are quite successful. First, we focus on the nuclear matter (NM) system, where the Coulomb and surface interactions are neglected. The binding energy per particle of the NM system is $\approx-16$ MeV at the saturation density $\rho_0\sim 0.148$ fm$^{-3}$ \cite{NKGb_1997}. The characteristics EOS of the NM is calculated by using different force interactions \cite{Dutra_2012}. There are  some  empirical/experimental data  to constraint the NM EOSs as given in Refs. \cite{Danielewicz_2002, Kumar_2018}. Different NM quantities such as incompressibility, symmetry energy and its slope parameter etc. play important role to explore the NS properties \cite{MCentelles_2009,Xu_2010,Fattoyev_2012,Steiner_2012,Newton_2012,Dutra_2012,Singh_2013}. In this study, our motivation is to constraint these NM parameters using recent experimental data \cite{Adhikari_2021,Reed_2021,Miller_2021}. 

Recently, the updated Lead Radius Experiment (PREX) has given the neutron skin thickness of $^{208}$Pb as $R_{{\rm skin}}=0.283\pm0.071$ fm \cite{Adhikari_2021}. Based on this data Patnaik {\it et al.} \cite{Pattnaik_2021} tuned the G3 and IOPB-I parameter sets. The impacts of PREX-2 data on the NM and NS properties have been explicitly studied in the Ref. \cite{Reed_2021}. The inferred values of NM quantities such as symmetry energy ($J$), its slope parameter ($L$) are $38.1\pm4.7$ MeV and $106\pm37$ MeV, respectively. The inferred limits are systematically larger as compared with either theoretical or experimental values \cite{ZHANG_2013,Horowitz_2001,Ducoin_2010,Ducoin_2011,Hebeler_2013,Horowitz_2014, Drischler_2020, Hagen_2016, Chen_2010, Steiner_2012,Gandolfi_2014,Roca_2015}. Reed {\it et al.} has also been calculated the NS properties by combining old NICER and PREX-2 constraints. The predicted radius range is $13.25<R_{1.4}<14.26$ km. Recently, NICER also put a revised limit, which is inferred by combining the old NICER data, $\sim 2 \ M_\odot$ pulsars, and the tidal deformability constraints form GWs data, and different EOSs modeling. The new NICER radius range for the canonical star is $12.45\pm0.65$ km. In this study, we want to constraint the radius of the canonical NS using both NICER and Reed {\it et al.} data. 

The paper is organized as follows: The formalism for the calculation of different NM quantities is given in Sec. \ref{form}. The results and discussions on the NM and NS properties are provided in Sec. \ref{form:RD}. A summary of our work is enumerated in the Sec. \ref{summ}.
%%%%%%
\section{Formalism}
\label{form}
The energy density   ${\cal{E}}(\rho,\alpha)$ of the NM system can be expanded in a Taylor series in terms of asymmetry factor $\alpha  \big(=\frac{\rho_n-\rho_p}{\rho_n+\rho_p}\big)$ \cite{Horowitz_2014,Kumar_2018}:
\begin{equation}
{\cal{E}}(\rho,\alpha) = {\cal{E}}(\rho)+S(\rho) \xi^2+ {\cal{O}}(\xi^4),
\label{dere}
\end{equation}
where $\rho$, $\rho_n$, and $\rho_p$ are the total baryon, neutron, and proton densities, respectively, The $ {\cal{E}}(\rho)$ is the energy density of the symmetric NM. The density dependence symmetry energy ($S(\rho)$) can be written as 
\begin{equation}
S(\rho) = \frac{1}{2}\Bigg(\frac{\partial^2{\cal{E}}}{\partial\alpha^2}\Bigg)_{\alpha=0}.
\label{sym1}
\end{equation}
The value of $S(\rho)$ is the most uncertain property of the NM, and it has a large diversion at a high-density limit \cite{BaoLi_2019}. Many progress have been made to constrain the $S(\rho)$ starting from heavy-ion collision to NS \cite{BaoLi_2013, Danielewicz_2002}. Here, we can expand the $S(\rho)$ in a leptodermous expansion near the saturation density  as follow \cite{Matsui_1981,Kubis_1997,MCentelles_2001,Chen_2014,Kumar_2018}:
\begin{equation}
S(\rho) = J+L\xi+\frac{1}{2}K_{sym}\xi^2+\frac{1}{6}Q_{sym}\xi^3+{\cal{O}}(\xi^4),
\label{eq11}
\end{equation}
where $\xi=\frac{\rho-\rho_0}{3\rho_0}$, $J$ is the symmetry energy at saturation density $\rho_0$ and the other parameters like slope ($L$), curvature ($K_{sym}$) and skewness ($Q_{sym}$) are given as follow:
\begin{eqnarray}
L=3\rho\frac{\partial S(\rho)}{\partial\rho}\Big|_{\rho=\rho_0},\\
K_{sym}=9\rho^2\frac{\partial^2 S(\rho)}{\partial\rho^2}\Big|_{\rho=\rho_0},\\
Q_{sym}=27\rho^3\frac{\partial^3 S(\rho)}{\partial\rho^3}\Big|_{\rho=\rho_0}.
\end{eqnarray}
%%%
Similarly, one can expand the asymmetric NM incompressiblity $K(\alpha)$ as 
\begin{equation}
K(\alpha)=K+K_\tau \alpha^2+{\cal{O}}(\alpha^4),
\end{equation}
%%%%%
where $K$ is the incompressibility of the NM at the saturation density and
\begin{equation}
K_\tau= K_{sym.}-6L-\frac{Q_0 L}{K},
\label{ktau}
\end{equation}
with $Q_0=27\rho^3\frac{\partial^3 \cal{E}}{\partial {\rho}^3}$ in symmetric NM at saturation density. We use another quantity $K^\prime=-Q_0$.
%%%%
\section{Results and Discussions}
\label{form:RD}
%%%%
\subsubsection{Nuclear Matter Properties}
\label{NM}
In this section, we constrain the values of $J$ and $L$. Moreover, with the addition of NM properties, we also want to constraint the mass and radius of the NS with the help of recent NICER data. To calculate NM properties we take 224 RMF, 240 SHF, 7 DD-RMF, 18 PC parameter sets from the Dutra {\it et al.} \cite{Dutra_2012, Dutra_2014}, which span a large parameter space. For the calculation of NS properties, we take 50 well-known EOSs from the Refs. \cite{Fortin_2016, Biswal_2020}.
%%%%%%%%%%%%%
\begin{figure}
\centering
\includegraphics[width=0.5\textwidth]{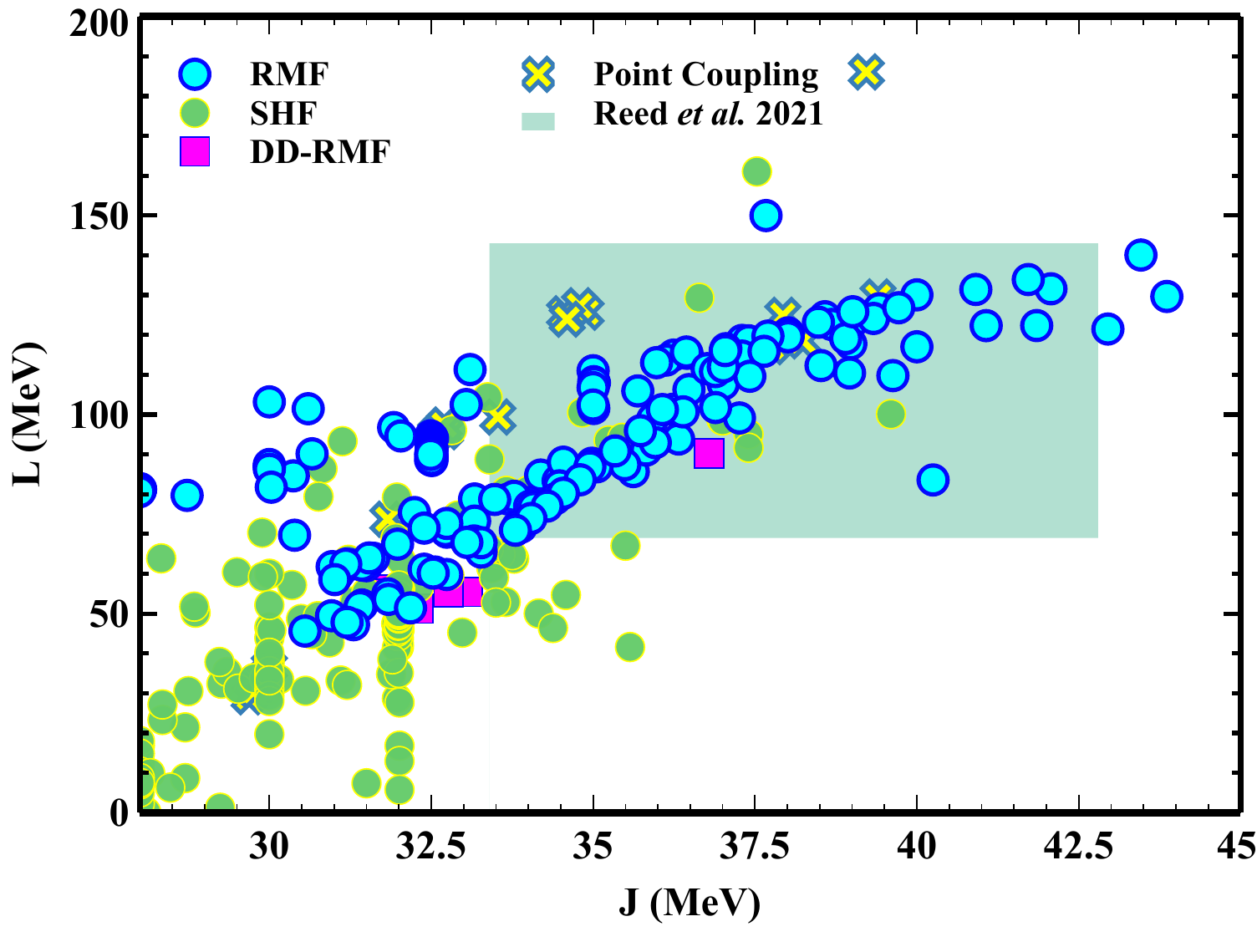}
\caption{(color online) The value of slope parameter for 224 RMF, 240 SHF, 7 DD-RMF and 18 PC parameter sets are shown. The light green box represents the values of $J=38.1\pm4.7$ MeV and $L=106\pm37$ MeV taken from the Ref. \cite{Reed_2021}. }
\label{fig:LS}
\end{figure}
%%%%%%%%%%%%%%%

In Fig. \ref{fig:LS}, we plot the symmetry energy (J) as a function of the slope parameter (L) for the considered parameter sets. Recently, the PREX-II experiments put a limit on the skin thickness of $^{208}$Pb is $R_{\mathrm{skin}}=0.283\pm0.071$ fm \cite{Adhikari_2021}. There is strong correlation have been observed between $R_{\mathrm{skin}}$ and $L$ at saturation density \cite{Roca_2011,Reed_2021}. Reed {\it et al.} \cite{Reed_2021} inferred the values of $J$ and $L$  as $38.1\pm4.7$ MeV and $106\pm37$ MeV respectively. We put a light green box in Fig. \ref{fig:LS} to constraint the values of $J$ and $L$. Most of the RMF and PC parameter sets satisfy  the constraints well as compare to SHF and DDRMF parameter sets. Hence, we conclude that the parameter sets which predict higher values of $J$ and $L$ are consistent with Reed {\it et al.} data.

\begin{figure}
\centering
\includegraphics[width=0.5\textwidth]{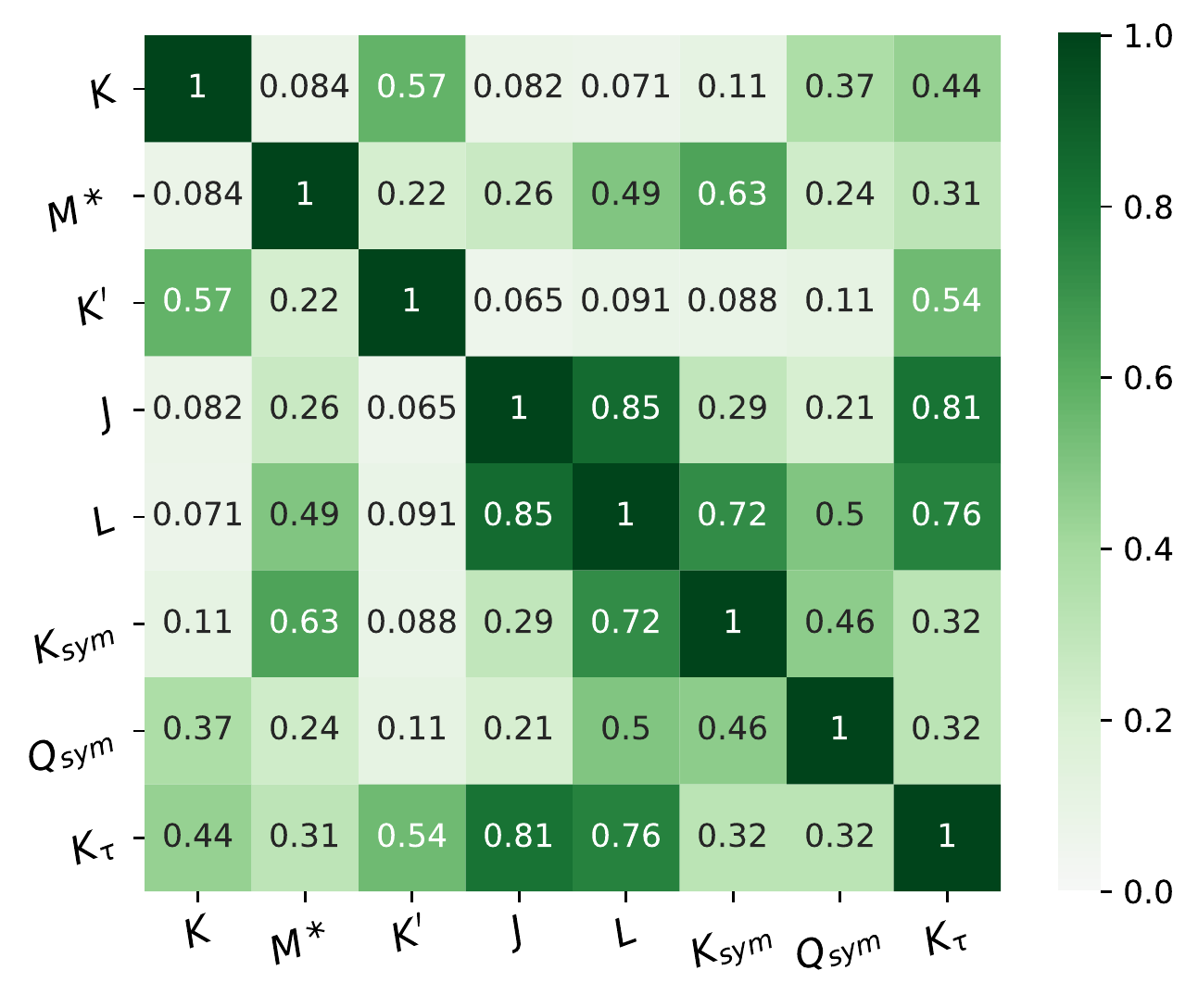}
\caption{(color online) Correlations matrix represent the correlation between different NM parameters. The number inside the box represents the correlation coefficient between corresponding parameters.}
\label{fig:corr}
\end{figure}

To check the correlation between different NM parameters as calculated in Sub-Sec. \ref{NM}, we plot the correlation matrix as shown in Fig. \ref{fig:corr}. To check the linear correlation between pairs of quantities, we calculate the correlation coefficient using Pearson's formula as used in Ref. \cite{Biswal_2020}. We find that the correlation coefficient between $J$ and $L$ is found to be 0.85. Slightly weaker correlations are found between $K_\tau$--$J$ and $K_\tau$--$L$ with coefficients 0.81 and 0.76, respectively.
%%%%%%%%%%%%%
%%%%%%%%%%%%%%%
\subsection{Neutron Star Properties}
\label{NS}
\begin{figure}
\centering
\includegraphics[width=0.5\textwidth]{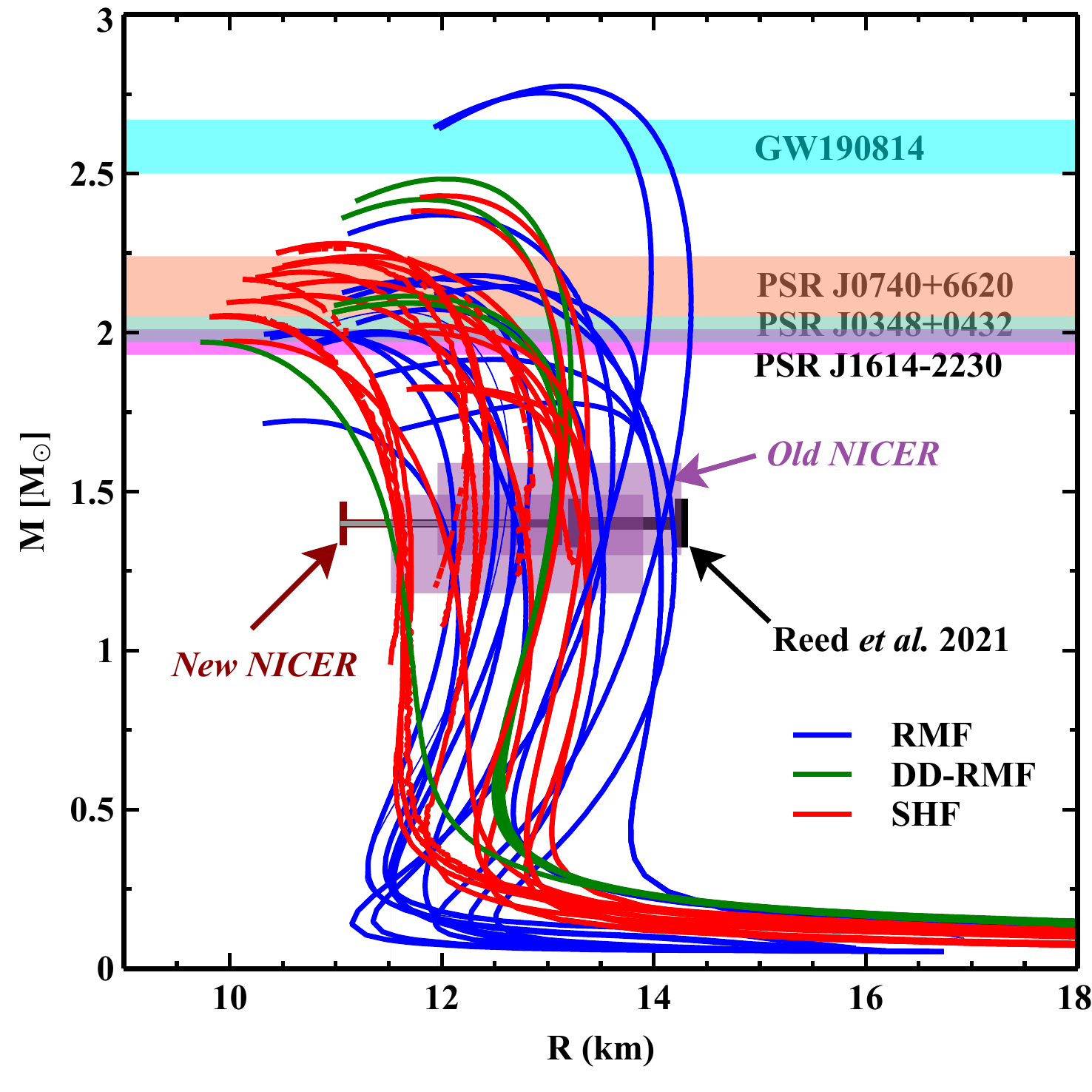}
\caption{(color online) $M$--$R$ are shown for RMF, SHF and DD-RMF sets with different observational constraints. Both old NICER data shown with two violet boxes \cite{Miller_2019, Riley_2019} and new NICER data with dark red doubled head lines are also depicted \cite{Miller_2021} for canonical star. Recently radius constraint given by Reed {\it et al.} \cite{Reed_2021} is also shown with black headed line. }
\label{fig:mr}
\end{figure}

The NS is composed of neutrons, protons, and leptons. Inside the NS, the neutron decays to proton, electron, and anti-neutrino. This process is called $\beta$-decay. Both $\beta$-equilibrium and charge neutrality processes are required for the stability  of the NS \cite{NKGb_1997}. Therefore, the total EOS of the NS is the addition of baryons and leptons as given as \cite{Kumar_2018, Das_2020}
%%%%%
\begin{eqnarray}
{\cal{E}}_{NS}=\mathcal{E}+\mathcal{E}_l, \ P_{NS}=P+P_l,
\end{eqnarray}
%%%%%
where $\mathcal{E} \ (\mathcal{E}_l)$ is the energy density of the NM (leptons) as given in Refs. \cite{Das_2020, Das_2021,das2021effects}. The $l$ correspond to both electron ($e^-$) and muon ($\mu^-$).
%%%%%%%%%%%%%
\begin{figure}
\centering
\includegraphics[width=0.5\textwidth]{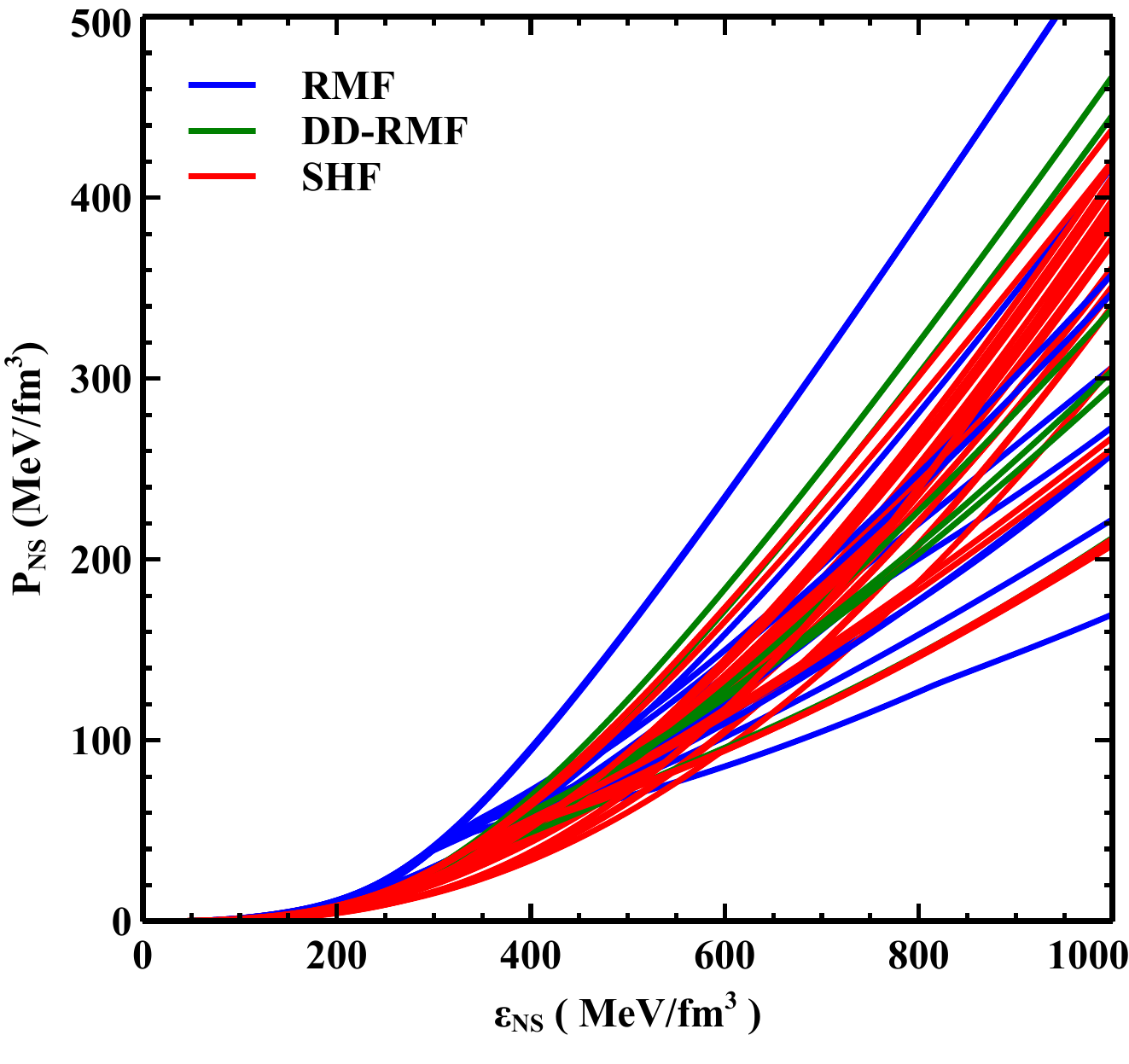}
\caption{(color online) EOSs are shown for RMF, SHF and DD-RMF sets.}
\label{fig:eos}
\end{figure}
%%%%%%%%%%%%%%%
  In Fig. \ref{fig:eos} , we plot  50 selected unified  EOS taken from the Refs. \cite{Fortin_2016, Biswal_2020} for comparison. The mass and radius of the NS are  calculated by solving Tolman-Oppenheimer-Volkoff equations \cite{TOV1, TOV2} with boundary conditions $P(0)=P_c$ and $P(R)=0$ for a fixed central density. Each EOS gives a different maximum mass and radius. We calculate the mass-radius ($M$-$R$) profile of the NS for considered EOSs as shown in Fig. \ref{fig:mr}. The massive pulsars data such as PSR J1614-2230 \cite{Demorest_2010}, PSR JO348+0432 \cite{Antoniadis_2013} and PSR J0740+6620 \cite{Cromartie_2019} are shown with different colour bars. Recently, the secondary component of the GW190814 event is observed in the mass range 2.50--2.67 $M_\odot$. A lot of debates are in progress, whether the secondary component is the lightest black hole or heaviest NS \cite{RAbbott_2020,DasPRD_2021}. The old NICER data depicted with two violet boxes from two different analyses \cite{Miller_2019, Riley_2019}. New NICER data \cite{Miller_2021} is shown with a double-headed dark red line. Recently, Reed {\it et al.} \cite{Reed_2021} have given radius constraints for canonical star inferred from the PREX-2, and old NICER data is also shown with double-headed black line.
%%%%%%%%%%%%%
%%%%%%

From the Fig. \ref{fig:mr}, it is clear that all EOSs reproduces $\sim 2 \ M_\odot$ except few RMF EOSs. None of the EOSs reproduces the GW190814 data. Almost all EOSs satisfy old NICER data, as clearly visible in the figure. The old NICER radius range is 11.52--14.26 km which provides a wider limit. But new NICER radius range is 11.8--13.1 km, which is a narrow band compared to old NICER data. Reed {\it et al.} radius range is 13.25--14.26 km which satisfy by some RMF and few SHF EOSs. If one carefully observe this three radius limits, there is a large uncertainty to constraint the radius of the NS. Future observation may put a tight constraint on the radius of the NS.   
%%%%%%
\section{Summary}
\label{summ}
In this manuscript, we calculate the NM properties with different formalisms such as RMF, SHF, DD-RMF, and PC. The symmetry energy and its slope are constrained by the recently inferred data from the PREX-2 experiment. We find that a numbers of RMF parameter sets, almost all PC  parameter sets, and few SHF and DDRMF  parameter sets satisfy the constraints given by Reed {\it et al.}. We also obtain some correlations between $J-L$, $K_\tau-J$ and $K_\tau-L$ with correlation coefficients 0.85, 0.81 and 0.76 respectively.

For the NS case, we take 50 unified EOSs and calculate the mass and radius. Almost all EOSs satisfy $2\ M_\odot$ constraints. None of the considered parameter sets satisfy the GW190814 data. The old NICER radius range is reproduced by almost all considered EOSs as it spans over a broader radius region. But the new NICER data is reproduced by almost all SHF and DD-RMF with few of the RMF sets. From the PREX-2 and old NICER data, Reed {\it et al.} inferred the radius of the canonical star, which is shown with a double-headed black line. Only stiff EOSs satisfy  the Reed {\it et al.} data because the radius range is a little bit higher as compare to new NICER data. 

In conclusion, the PREX-2 data supports the EOSs, which are stiffer. Therefore, all the NM and NS properties inferred from PREX-2 data provide large value symmetry energy coefficient as compare to other theoretical/experimental data. In the case of the NS, the implied radius range is also push towards a higher value, which can only be  supported  by stiffer EOSs. But new NICER data provides the radius range, which supports by the softer EOSs. Therefore, a possible uncertainty  has been developed to constraints the radius of the NS. We hope future experiments/observations may answer this question. 
%%%%%
\bibliography{NM}

%merlin.mbs apsrev4-1.bst 2010-07-25 4.21a (PWD, AO, DPC) hacked
%Control: key (0)
%Control: author (72) initials jnrlst
%Control: editor formatted (1) identically to author
%Control: production of article title (-1) disabled
%Control: page (0) single
%Control: year (1) truncated
%Control: production of eprint (0) enabled
\begin{thebibliography}{73}%
\makeatletter
\providecommand \@ifxundefined [1]{%
 \@ifx{#1\undefined}
}%
\providecommand \@ifnum [1]{%
 \ifnum #1\expandafter \@firstoftwo
 \else \expandafter \@secondoftwo
 \fi
}%
\providecommand \@ifx [1]{%
 \ifx #1\expandafter \@firstoftwo
 \else \expandafter \@secondoftwo
 \fi
}%
\providecommand \natexlab [1]{#1}%
\providecommand \enquote  [1]{``#1''}%
\providecommand \bibnamefont  [1]{#1}%
\providecommand \bibfnamefont [1]{#1}%
\providecommand \citenamefont [1]{#1}%
\providecommand \href@noop [0]{\@secondoftwo}%
\providecommand \href [0]{\begingroup \@sanitize@url \@href}%
\providecommand \@href[1]{\@@startlink{#1}\@@href}%
\providecommand \@@href[1]{\endgroup#1\@@endlink}%
\providecommand \@sanitize@url [0]{\catcode `\\12\catcode `\$12\catcode
  `\&12\catcode `\#12\catcode `\^12\catcode `\_12\catcode `\%12\relax}%
\providecommand \@@startlink[1]{}%
\providecommand \@@endlink[0]{}%
\providecommand \url  [0]{\begingroup\@sanitize@url \@url }%
\providecommand \@url [1]{\endgroup\@href {#1}{\urlprefix }}%
\providecommand \urlprefix  [0]{URL }%
\providecommand \Eprint [0]{\href }%
\providecommand \doibase [0]{http://dx.doi.org/}%
\providecommand \selectlanguage [0]{\@gobble}%
\providecommand \bibinfo  [0]{\@secondoftwo}%
\providecommand \bibfield  [0]{\@secondoftwo}%
\providecommand \translation [1]{[#1]}%
\providecommand \BibitemOpen [0]{}%
\providecommand \bibitemStop [0]{}%
\providecommand \bibitemNoStop [0]{.\EOS\space}%
\providecommand \EOS [0]{\spacefactor3000\relax}%
\providecommand \BibitemShut  [1]{\csname bibitem#1\endcsname}%
\let\auto@bib@innerbib\@empty
%</preamble>
\bibitem [{\citenamefont {Lattimer}\ and\ \citenamefont
  {Prakash}(2004)}]{Lattimer_2004}%
  \BibitemOpen
  \bibfield  {author} {\bibinfo {author} {\bibfnamefont {J.~M.}\ \bibnamefont
  {Lattimer}}\ and\ \bibinfo {author} {\bibfnamefont {M.}~\bibnamefont
  {Prakash}},\ }\href {\doibase 10.1126/science.1090720} {\bibfield  {journal}
  {\bibinfo  {journal} {Science}\ }\textbf {\bibinfo {volume} {304}},\ \bibinfo
  {pages} {536} (\bibinfo {year} {2004})}\BibitemShut {NoStop}%
\bibitem [{\citenamefont {{Ambartsumyan}}\ and\ \citenamefont
  {{Saakyan}}(1960)}]{Ambartsumyan_1960}%
  \BibitemOpen
  \bibfield  {author} {\bibinfo {author} {\bibfnamefont {V.~A.}\ \bibnamefont
  {{Ambartsumyan}}}\ and\ \bibinfo {author} {\bibfnamefont {G.~S.}\
  \bibnamefont {{Saakyan}}},\ }\href@noop {} {\bibfield  {journal} {\bibinfo
  {journal} {sovast}\ }\textbf {\bibinfo {volume} {4}},\ \bibinfo {pages} {187}
  (\bibinfo {year} {1960})}\BibitemShut {NoStop}%
\bibitem [{\citenamefont {{Glendenning}}(1985)}]{NKGh_1985}%
  \BibitemOpen
  \bibfield  {author} {\bibinfo {author} {\bibfnamefont {N.~K.}\ \bibnamefont
  {{Glendenning}}},\ }\href {\doibase 10.1086/163253} {\bibfield  {journal}
  {\bibinfo  {journal} {APJ}\ }\textbf {\bibinfo {volume} {293}},\ \bibinfo
  {pages} {470} (\bibinfo {year} {1985})}\BibitemShut {NoStop}%
\bibitem [{\citenamefont {Schaffner}\ and\ \citenamefont
  {Mishustin}(1996)}]{Schaffner_1996}%
  \BibitemOpen
  \bibfield  {author} {\bibinfo {author} {\bibfnamefont {J.}~\bibnamefont
  {Schaffner}}\ and\ \bibinfo {author} {\bibfnamefont {I.~N.}\ \bibnamefont
  {Mishustin}},\ }\href {\doibase 10.1103/PhysRevC.53.1416} {\bibfield
  {journal} {\bibinfo  {journal} {Phys. Rev. C}\ }\textbf {\bibinfo {volume}
  {53}},\ \bibinfo {pages} {1416} (\bibinfo {year} {1996})}\BibitemShut
  {NoStop}%
\bibitem [{\citenamefont {Biswal}\ \emph {et~al.}(2016)\citenamefont {Biswal},
  \citenamefont {Kumar},\ and\ \citenamefont {Patra}}]{Biswal_2016}%
  \BibitemOpen
  \bibfield  {author} {\bibinfo {author} {\bibfnamefont {S.~K.}\ \bibnamefont
  {Biswal}}, \bibinfo {author} {\bibfnamefont {B.}~\bibnamefont {Kumar}}, \
  and\ \bibinfo {author} {\bibfnamefont {S.~K.}\ \bibnamefont {Patra}},\
  }\href@noop {} {} (\bibinfo {year} {2016}),\ \Eprint
  {http://arxiv.org/abs/1602.08888} {arXiv:1602.08888 [nucl-th]} \BibitemShut
  {NoStop}%
\bibitem [{\citenamefont {Fortin}\ \emph {et~al.}(2017)\citenamefont {Fortin},
  \citenamefont {Avancini}, \citenamefont {Provid\^encia},\ and\ \citenamefont
  {Vida\~na}}]{Fortin_2017}%
  \BibitemOpen
  \bibfield  {author} {\bibinfo {author} {\bibfnamefont {M.}~\bibnamefont
  {Fortin}}, \bibinfo {author} {\bibfnamefont {S.~S.}\ \bibnamefont
  {Avancini}}, \bibinfo {author} {\bibfnamefont {C.}~\bibnamefont
  {Provid\^encia}}, \ and\ \bibinfo {author} {\bibfnamefont {I.}~\bibnamefont
  {Vida\~na}},\ }\href {\doibase 10.1103/PhysRevC.95.065803} {\bibfield
  {journal} {\bibinfo  {journal} {Phys. Rev. C}\ }\textbf {\bibinfo {volume}
  {95}},\ \bibinfo {pages} {065803} (\bibinfo {year} {2017})}\BibitemShut
  {NoStop}%
\bibitem [{\citenamefont {Bhuyan}\ \emph {et~al.}(2017)\citenamefont {Bhuyan},
  \citenamefont {Carlson}, \citenamefont {Patra},\ and\ \citenamefont
  {Zhou}}]{Bhuyan_2017}%
  \BibitemOpen
  \bibfield  {author} {\bibinfo {author} {\bibfnamefont {M.}~\bibnamefont
  {Bhuyan}}, \bibinfo {author} {\bibfnamefont {B.~V.}\ \bibnamefont {Carlson}},
  \bibinfo {author} {\bibfnamefont {S.~K.}\ \bibnamefont {Patra}}, \ and\
  \bibinfo {author} {\bibfnamefont {S.-G.}\ \bibnamefont {Zhou}},\ }\href
  {\doibase 10.1142/s0218301317500525} {\bibfield  {journal} {\bibinfo
  {journal} {IJMP E}\ }\textbf {\bibinfo {volume} {26}},\ \bibinfo {pages}
  {1750052} (\bibinfo {year} {2017})}\BibitemShut {NoStop}%
\bibitem [{\citenamefont {Biswal}(2019)}]{Biswalaip_2019}%
  \BibitemOpen
  \bibfield  {author} {\bibinfo {author} {\bibfnamefont {S.~K.}\ \bibnamefont
  {Biswal}},\ }\href {\doibase 10.1063/1.5117821} {\bibfield  {journal}
  {\bibinfo  {journal} {AIP Conf. Proceedings}\ }\textbf {\bibinfo {volume}
  {2127}},\ \bibinfo {pages} {020031} (\bibinfo {year} {2019})}\BibitemShut
  {NoStop}%
\bibitem [{\citenamefont {Biswal}\ \emph {et~al.}(2019)\citenamefont {Biswal},
  \citenamefont {Patra},\ and\ \citenamefont {Zhou}}]{Biswal_2019}%
  \BibitemOpen
  \bibfield  {author} {\bibinfo {author} {\bibfnamefont {S.~K.}\ \bibnamefont
  {Biswal}}, \bibinfo {author} {\bibfnamefont {S.~K.}\ \bibnamefont {Patra}}, \
  and\ \bibinfo {author} {\bibfnamefont {S.-G.}\ \bibnamefont {Zhou}},\ }\href
  {\doibase 10.3847/1538-4357/ab43c5} {\bibfield  {journal} {\bibinfo
  {journal} {APJ}\ }\textbf {\bibinfo {volume} {885}},\ \bibinfo {pages} {25}
  (\bibinfo {year} {2019})}\BibitemShut {NoStop}%
\bibitem [{\citenamefont {Collins}\ and\ \citenamefont
  {Perry}(1975)}]{Collins_1975}%
  \BibitemOpen
  \bibfield  {author} {\bibinfo {author} {\bibfnamefont {J.~C.}\ \bibnamefont
  {Collins}}\ and\ \bibinfo {author} {\bibfnamefont {M.~J.}\ \bibnamefont
  {Perry}},\ }\href {\doibase 10.1103/PhysRevLett.34.1353} {\bibfield
  {journal} {\bibinfo  {journal} {Phys. Rev. Lett.}\ }\textbf {\bibinfo
  {volume} {34}},\ \bibinfo {pages} {1353} (\bibinfo {year}
  {1975})}\BibitemShut {NoStop}%
\bibitem [{\citenamefont {Orsaria}\ \emph {et~al.}(2014)\citenamefont
  {Orsaria}, \citenamefont {Rodrigues}, \citenamefont {Weber},\ and\
  \citenamefont {Contrera}}]{Orsaria_2014}%
  \BibitemOpen
  \bibfield  {author} {\bibinfo {author} {\bibfnamefont {M.}~\bibnamefont
  {Orsaria}}, \bibinfo {author} {\bibfnamefont {H.}~\bibnamefont {Rodrigues}},
  \bibinfo {author} {\bibfnamefont {F.}~\bibnamefont {Weber}}, \ and\ \bibinfo
  {author} {\bibfnamefont {G.~A.}\ \bibnamefont {Contrera}},\ }\href {\doibase
  10.1103/PhysRevC.89.015806} {\bibfield  {journal} {\bibinfo  {journal} {Phys.
  Rev. C}\ }\textbf {\bibinfo {volume} {89}},\ \bibinfo {pages} {015806}
  (\bibinfo {year} {2014})}\BibitemShut {NoStop}%
\bibitem [{\citenamefont {Mellinger}\ \emph {et~al.}(2017)\citenamefont
  {Mellinger}, \citenamefont {Weber}, \citenamefont {Spinella}, \citenamefont
  {Contrera},\ and\ \citenamefont {Orsaria}}]{Mellinger_2017}%
  \BibitemOpen
  \bibfield  {author} {\bibinfo {author} {\bibfnamefont {R.}~\bibnamefont
  {Mellinger}}, \bibinfo {author} {\bibfnamefont {F.}~\bibnamefont {Weber}},
  \bibinfo {author} {\bibfnamefont {W.}~\bibnamefont {Spinella}}, \bibinfo
  {author} {\bibfnamefont {G.}~\bibnamefont {Contrera}}, \ and\ \bibinfo
  {author} {\bibfnamefont {M.}~\bibnamefont {Orsaria}},\ }\href {\doibase
  10.3390/universe3010005} {\bibfield  {journal} {\bibinfo  {journal}
  {Universe}\ }\textbf {\bibinfo {volume} {3}},\ \bibinfo {pages} {5} (\bibinfo
  {year} {2017})}\BibitemShut {NoStop}%
\bibitem [{\citenamefont {Miller}\ and\ \citenamefont
  {Green}(1972)}]{Miller_1972}%
  \BibitemOpen
  \bibfield  {author} {\bibinfo {author} {\bibfnamefont {L.~D.}\ \bibnamefont
  {Miller}}\ and\ \bibinfo {author} {\bibfnamefont {A.~E.~S.}\ \bibnamefont
  {Green}},\ }\href {\doibase 10.1103/PhysRevC.5.241} {\bibfield  {journal}
  {\bibinfo  {journal} {Phys. Rev. C}\ }\textbf {\bibinfo {volume} {5}},\
  \bibinfo {pages} {241} (\bibinfo {year} {1972})}\BibitemShut {NoStop}%
\bibitem [{\citenamefont {Serot}\ and\ \citenamefont
  {Walecka}(1986)}]{Serot_1986}%
  \BibitemOpen
  \bibfield  {author} {\bibinfo {author} {\bibfnamefont {B.~D.}\ \bibnamefont
  {Serot}}\ and\ \bibinfo {author} {\bibfnamefont {J.~D.}\ \bibnamefont
  {Walecka}},\ }\href@noop {} {\bibfield  {journal} {\bibinfo  {journal} {Adv.
  Nucl. Phys.}\ }\textbf {\bibinfo {volume} {16}},\ \bibinfo {pages} {1}
  (\bibinfo {year} {1986})}\BibitemShut {NoStop}%
%%CITATION = ANUPB,16,1;%%
\bibitem [{\citenamefont {Furnstahl}\ \emph {et~al.}(1987)\citenamefont
  {Furnstahl}, \citenamefont {Price},\ and\ \citenamefont
  {Walker}}]{Furn_1987}%
  \BibitemOpen
  \bibfield  {author} {\bibinfo {author} {\bibfnamefont {R.~J.}\ \bibnamefont
  {Furnstahl}}, \bibinfo {author} {\bibfnamefont {C.~E.}\ \bibnamefont
  {Price}}, \ and\ \bibinfo {author} {\bibfnamefont {G.~E.}\ \bibnamefont
  {Walker}},\ }\href {\doibase 10.1103/PhysRevC.36.2590} {\bibfield  {journal}
  {\bibinfo  {journal} {Phys. Rev. C}\ }\textbf {\bibinfo {volume} {36}},\
  \bibinfo {pages} {2590} (\bibinfo {year} {1987})}\BibitemShut {NoStop}%
\bibitem [{\citenamefont {Reinhard}(1988)}]{Reinhard_1988}%
  \BibitemOpen
  \bibfield  {author} {\bibinfo {author} {\bibfnamefont {P.~G.}\ \bibnamefont
  {Reinhard}},\ }\href {\doibase 10.1007/BF01290231} {\bibfield  {journal}
  {\bibinfo  {journal} {Z. Phys. A Atomic Nuclei}\ }\textbf {\bibinfo {volume}
  {329}},\ \bibinfo {pages} {257} (\bibinfo {year} {1988})}\BibitemShut
  {NoStop}%
\bibitem [{\citenamefont {Furnstahl}\ \emph {et~al.}(1997)\citenamefont
  {Furnstahl}, \citenamefont {Serot},\ and\ \citenamefont {Tang}}]{Frun_1997}%
  \BibitemOpen
  \bibfield  {author} {\bibinfo {author} {\bibfnamefont {R.~J.}\ \bibnamefont
  {Furnstahl}}, \bibinfo {author} {\bibfnamefont {B.~D.}\ \bibnamefont
  {Serot}}, \ and\ \bibinfo {author} {\bibfnamefont {H.-B.}\ \bibnamefont
  {Tang}},\ }\href {\doibase https://doi.org/10.1016/S0375-9474(96)00472-1}
  {\bibfield  {journal} {\bibinfo  {journal} {Nucl. Phys. A}\ }\textbf
  {\bibinfo {volume} {615}},\ \bibinfo {pages} {441 } (\bibinfo {year}
  {1997})},\ \Eprint {http://arxiv.org/abs/9608035} {arXiv:9608035 [nucl-th]}
  \BibitemShut {NoStop}%
\bibitem [{\citenamefont {Kumar}\ \emph {et~al.}(2017)\citenamefont {Kumar},
  \citenamefont {Singh}, \citenamefont {Agrawal},\ and\ \citenamefont
  {Patra}}]{Kumar_2017}%
  \BibitemOpen
  \bibfield  {author} {\bibinfo {author} {\bibfnamefont {B.}~\bibnamefont
  {Kumar}}, \bibinfo {author} {\bibfnamefont {S.}~\bibnamefont {Singh}},
  \bibinfo {author} {\bibfnamefont {B.}~\bibnamefont {Agrawal}}, \ and\
  \bibinfo {author} {\bibfnamefont {S.}~\bibnamefont {Patra}},\ }\href
  {\doibase https://doi.org/10.1016/j.nuclphysa.2017.07.001} {\bibfield
  {journal} {\bibinfo  {journal} {Nuclear Physics A}\ }\textbf {\bibinfo
  {volume} {966}},\ \bibinfo {pages} {197 } (\bibinfo {year}
  {2017})}\BibitemShut {NoStop}%
\bibitem [{\citenamefont {Kumar}\ \emph {et~al.}(2018)\citenamefont {Kumar},
  \citenamefont {Patra},\ and\ \citenamefont {Agrawal}}]{Kumar_2018}%
  \BibitemOpen
  \bibfield  {author} {\bibinfo {author} {\bibfnamefont {B.}~\bibnamefont
  {Kumar}}, \bibinfo {author} {\bibfnamefont {S.~K.}\ \bibnamefont {Patra}}, \
  and\ \bibinfo {author} {\bibfnamefont {B.~K.}\ \bibnamefont {Agrawal}},\
  }\href {\doibase 10.1103/PhysRevC.97.045806} {\bibfield  {journal} {\bibinfo
  {journal} {Phys. Rev. C}\ }\textbf {\bibinfo {volume} {97}},\ \bibinfo
  {pages} {045806} (\bibinfo {year} {2018})}\BibitemShut {NoStop}%
\bibitem [{\citenamefont {Skyrme}(1956)}]{Skyrme_1956}%
  \BibitemOpen
  \bibfield  {author} {\bibinfo {author} {\bibfnamefont {T.~H.~R.}\
  \bibnamefont {Skyrme}},\ }\href {\doibase 10.1080/14786435608238186}
  {\bibfield  {journal} {\bibinfo  {journal} {The Philosophical Magazine: A
  Journal of Theoretical Experimental and Applied Physics}\ }\textbf {\bibinfo
  {volume} {1}},\ \bibinfo {pages} {1043} (\bibinfo {year} {1956})}\BibitemShut
  {NoStop}%
\bibitem [{\citenamefont {Skyrme}(1958)}]{Skyrme_1958}%
  \BibitemOpen
  \bibfield  {author} {\bibinfo {author} {\bibfnamefont {T.}~\bibnamefont
  {Skyrme}},\ }\href {\doibase https://doi.org/10.1016/0029-5582(58)90345-6}
  {\bibfield  {journal} {\bibinfo  {journal} {Nuclear Physics}\ }\textbf
  {\bibinfo {volume} {9}},\ \bibinfo {pages} {615 } (\bibinfo {year}
  {1958})}\BibitemShut {NoStop}%
\bibitem [{\citenamefont {Vautherin}\ and\ \citenamefont
  {Brink}(1972)}]{Vautherian_1972}%
  \BibitemOpen
  \bibfield  {author} {\bibinfo {author} {\bibfnamefont {D.}~\bibnamefont
  {Vautherin}}\ and\ \bibinfo {author} {\bibfnamefont {D.~M.}\ \bibnamefont
  {Brink}},\ }\href {\doibase 10.1103/PhysRevC.5.626} {\bibfield  {journal}
  {\bibinfo  {journal} {Phys. Rev. C}\ }\textbf {\bibinfo {volume} {5}},\
  \bibinfo {pages} {626} (\bibinfo {year} {1972})}\BibitemShut {NoStop}%
\bibitem [{\citenamefont {Chabanat}\ \emph {et~al.}(1998)\citenamefont
  {Chabanat}, \citenamefont {Bonche}, \citenamefont {Haensel}, \citenamefont
  {Meyer},\ and\ \citenamefont {Schaeffer"}}]{Chabanta_1998}%
  \BibitemOpen
  \bibfield  {author} {\bibinfo {author} {\bibfnamefont {E.}~\bibnamefont
  {Chabanat}}, \bibinfo {author} {\bibfnamefont {P.}~\bibnamefont {Bonche}},
  \bibinfo {author} {\bibfnamefont {P.}~\bibnamefont {Haensel}}, \bibinfo
  {author} {\bibfnamefont {J.}~\bibnamefont {Meyer}}, \ and\ \bibinfo {author}
  {\bibfnamefont {R.}~\bibnamefont {Schaeffer"}},\ }\href {\doibase
  https://doi.org/10.1016/S0375-9474(98)00180-8} {\bibfield  {journal}
  {\bibinfo  {journal} {Nucl. Phys. A}\ }\textbf {\bibinfo {volume} {635}},\
  \bibinfo {pages} {231 } (\bibinfo {year} {1998})}\BibitemShut {NoStop}%
\bibitem [{\citenamefont {Alex~Brown}(1998)}]{Brown_1998}%
  \BibitemOpen
  \bibfield  {author} {\bibinfo {author} {\bibfnamefont {B.}~\bibnamefont
  {Alex~Brown}},\ }\href {\doibase 10.1103/PhysRevC.58.220} {\bibfield
  {journal} {\bibinfo  {journal} {Phys. Rev. C}\ }\textbf {\bibinfo {volume}
  {58}},\ \bibinfo {pages} {220} (\bibinfo {year} {1998})}\BibitemShut
  {NoStop}%
\bibitem [{\citenamefont {Stone}\ and\ \citenamefont
  {Reinhard}(2007)}]{Stone_2007}%
  \BibitemOpen
  \bibfield  {author} {\bibinfo {author} {\bibfnamefont {J.}~\bibnamefont
  {Stone}}\ and\ \bibinfo {author} {\bibfnamefont {P.-G.}\ \bibnamefont
  {Reinhard}},\ }\href {\doibase 10.1016/j.ppnp.2006.07.001} {\bibfield
  {journal} {\bibinfo  {journal} {Progress in Particle and Nuclear Physics}\
  }\textbf {\bibinfo {volume} {58}},\ \bibinfo {pages} {587–657} (\bibinfo
  {year} {2007})},\ \Eprint {http://arxiv.org/abs/nucl-th/0607002}
  {arXiv:nucl-th/0607002 [nucl-th]} \BibitemShut {NoStop}%
\bibitem [{\citenamefont {Dutra}\ \emph {et~al.}(2012)\citenamefont {Dutra},
  \citenamefont {Louren\ifmmode~\mbox{\c{c}}\else \c{c}\fi{}o}, \citenamefont
  {S\'a~Martins} \emph {et~al.}}]{Dutra_2012}%
  \BibitemOpen
  \bibfield  {author} {\bibinfo {author} {\bibfnamefont {M.}~\bibnamefont
  {Dutra}}, \bibinfo {author} {\bibfnamefont {O.}~\bibnamefont
  {Louren\ifmmode~\mbox{\c{c}}\else \c{c}\fi{}o}}, \bibinfo {author}
  {\bibfnamefont {J.~S.}\ \bibnamefont {S\'a~Martins}},  \emph {et~al.},\
  }\href {\doibase 10.1103/PhysRevC.85.035201} {\bibfield  {journal} {\bibinfo
  {journal} {Phys. Rev. C}\ }\textbf {\bibinfo {volume} {85}},\ \bibinfo
  {pages} {035201} (\bibinfo {year} {2012})}\BibitemShut {NoStop}%
\bibitem [{\citenamefont {Decharg\'e}\ and\ \citenamefont
  {Gogny}(1980)}]{Gogny_1980}%
  \BibitemOpen
  \bibfield  {author} {\bibinfo {author} {\bibfnamefont {J.}~\bibnamefont
  {Decharg\'e}}\ and\ \bibinfo {author} {\bibfnamefont {D.}~\bibnamefont
  {Gogny}},\ }\href {\doibase 10.1103/PhysRevC.21.1568} {\bibfield  {journal}
  {\bibinfo  {journal} {Phys. Rev. C}\ }\textbf {\bibinfo {volume} {21}},\
  \bibinfo {pages} {1568} (\bibinfo {year} {1980})}\BibitemShut {NoStop}%
\bibitem [{\citenamefont {Typel}(2005)}]{Typel_2005}%
  \BibitemOpen
  \bibfield  {author} {\bibinfo {author} {\bibfnamefont {S.}~\bibnamefont
  {Typel}},\ }\href {\doibase 10.1103/PhysRevC.71.064301} {\bibfield  {journal}
  {\bibinfo  {journal} {Phys. Rev. C}\ }\textbf {\bibinfo {volume} {71}},\
  \bibinfo {pages} {064301} (\bibinfo {year} {2005})}\BibitemShut {NoStop}%
\bibitem [{\citenamefont {Dutra}\ \emph {et~al.}(2014)\citenamefont {Dutra},
  \citenamefont {Louren\ifmmode~\mbox{\c{c}}\else \c{c}\fi{}o}, \citenamefont
  {Avancini} \emph {et~al.}}]{Dutra_2014}%
  \BibitemOpen
  \bibfield  {author} {\bibinfo {author} {\bibfnamefont {M.}~\bibnamefont
  {Dutra}}, \bibinfo {author} {\bibfnamefont {O.}~\bibnamefont
  {Louren\ifmmode~\mbox{\c{c}}\else \c{c}\fi{}o}}, \bibinfo {author}
  {\bibfnamefont {S.~S.}\ \bibnamefont {Avancini}},  \emph {et~al.},\ }\href
  {\doibase 10.1103/PhysRevC.90.055203} {\bibfield  {journal} {\bibinfo
  {journal} {Phys. Rev. C}\ }\textbf {\bibinfo {volume} {90}},\ \bibinfo
  {pages} {055203} (\bibinfo {year} {2014})}\BibitemShut {NoStop}%
\bibitem [{\citenamefont {Glendenning}(1997)}]{NKGb_1997}%
  \BibitemOpen
  \bibfield  {author} {\bibinfo {author} {\bibfnamefont {N.~K.}\ \bibnamefont
  {Glendenning}},\ }\href@noop {} {\emph {\bibinfo {title} {{Compact stars:
  Nuclear physics, particle physics, and general relativity}}}}\ (\bibinfo
  {publisher} {Springer-Verlag New York},\ \bibinfo {year} {1997})\BibitemShut
  {NoStop}%
%%CITATION = INSPIRE-456851;%%
\bibitem [{\citenamefont {Danielewicz}\ \emph {et~al.}(2002)\citenamefont
  {Danielewicz}, \citenamefont {Lacey},\ and\ \citenamefont
  {Lynch}}]{Danielewicz_2002}%
  \BibitemOpen
  \bibfield  {author} {\bibinfo {author} {\bibfnamefont {P.}~\bibnamefont
  {Danielewicz}}, \bibinfo {author} {\bibfnamefont {R.}~\bibnamefont {Lacey}},
  \ and\ \bibinfo {author} {\bibfnamefont {W.~G.}\ \bibnamefont {Lynch}},\
  }\href {\doibase 10.1126/science.1078070} {\bibfield  {journal} {\bibinfo
  {journal} {Science}\ }\textbf {\bibinfo {volume} {298}},\ \bibinfo {pages}
  {1592} (\bibinfo {year} {2002})}\BibitemShut {NoStop}%
\bibitem [{\citenamefont {Centelles}\ \emph {et~al.}(2009)\citenamefont
  {Centelles}, \citenamefont {Roca-Maza}, \citenamefont {Vi\~nas},\ and\
  \citenamefont {Warda}}]{MCentelles_2009}%
  \BibitemOpen
  \bibfield  {author} {\bibinfo {author} {\bibfnamefont {M.}~\bibnamefont
  {Centelles}}, \bibinfo {author} {\bibfnamefont {X.}~\bibnamefont
  {Roca-Maza}}, \bibinfo {author} {\bibfnamefont {X.}~\bibnamefont {Vi\~nas}},
  \ and\ \bibinfo {author} {\bibfnamefont {M.}~\bibnamefont {Warda}},\ }\href
  {\doibase 10.1103/PhysRevLett.102.122502} {\bibfield  {journal} {\bibinfo
  {journal} {Phys. Rev. Lett.}\ }\textbf {\bibinfo {volume} {102}},\ \bibinfo
  {pages} {122502} (\bibinfo {year} {2009})}\BibitemShut {NoStop}%
\bibitem [{\citenamefont {Xu}\ \emph {et~al.}(2010)\citenamefont {Xu},
  \citenamefont {Li},\ and\ \citenamefont {Chen}}]{Xu_2010}%
  \BibitemOpen
  \bibfield  {author} {\bibinfo {author} {\bibfnamefont {C.}~\bibnamefont
  {Xu}}, \bibinfo {author} {\bibfnamefont {B.-A.}\ \bibnamefont {Li}}, \ and\
  \bibinfo {author} {\bibfnamefont {L.-W.}\ \bibnamefont {Chen}},\ }\href
  {\doibase 10.1103/PhysRevC.82.054607} {\bibfield  {journal} {\bibinfo
  {journal} {Phys. Rev. C}\ }\textbf {\bibinfo {volume} {82}},\ \bibinfo
  {pages} {054607} (\bibinfo {year} {2010})}\BibitemShut {NoStop}%
\bibitem [{\citenamefont {Fattoyev}\ \emph {et~al.}(2012)\citenamefont
  {Fattoyev}, \citenamefont {Newton}, \citenamefont {Xu},\ and\ \citenamefont
  {Li}}]{Fattoyev_2012}%
  \BibitemOpen
  \bibfield  {author} {\bibinfo {author} {\bibfnamefont {F.~J.}\ \bibnamefont
  {Fattoyev}}, \bibinfo {author} {\bibfnamefont {W.~G.}\ \bibnamefont
  {Newton}}, \bibinfo {author} {\bibfnamefont {J.}~\bibnamefont {Xu}}, \ and\
  \bibinfo {author} {\bibfnamefont {B.-A.}\ \bibnamefont {Li}},\ }\href
  {\doibase 10.1103/PhysRevC.86.025804} {\bibfield  {journal} {\bibinfo
  {journal} {Phys. Rev. C}\ }\textbf {\bibinfo {volume} {86}},\ \bibinfo
  {pages} {025804} (\bibinfo {year} {2012})}\BibitemShut {NoStop}%
\bibitem [{\citenamefont {Steiner}\ and\ \citenamefont
  {Gandolfi}(2012)}]{Steiner_2012}%
  \BibitemOpen
  \bibfield  {author} {\bibinfo {author} {\bibfnamefont {A.~W.}\ \bibnamefont
  {Steiner}}\ and\ \bibinfo {author} {\bibfnamefont {S.}~\bibnamefont
  {Gandolfi}},\ }\href {\doibase 10.1103/PhysRevLett.108.081102} {\bibfield
  {journal} {\bibinfo  {journal} {Phys. Rev. Lett.}\ }\textbf {\bibinfo
  {volume} {108}},\ \bibinfo {pages} {081102} (\bibinfo {year}
  {2012})}\BibitemShut {NoStop}%
\bibitem [{\citenamefont {Newton}\ \emph {et~al.}(2012)\citenamefont {Newton},
  \citenamefont {Gearheart},\ and\ \citenamefont {Li}}]{Newton_2012}%
  \BibitemOpen
  \bibfield  {author} {\bibinfo {author} {\bibfnamefont {W.~G.}\ \bibnamefont
  {Newton}}, \bibinfo {author} {\bibfnamefont {M.}~\bibnamefont {Gearheart}}, \
  and\ \bibinfo {author} {\bibfnamefont {B.-A.}\ \bibnamefont {Li}},\ }\href
  {\doibase 10.1088/0067-0049/204/1/9} {\bibfield  {journal} {\bibinfo
  {journal} {APJ Suppl. Series}\ }\textbf {\bibinfo {volume} {204}},\ \bibinfo
  {pages} {9} (\bibinfo {year} {2012})}\BibitemShut {NoStop}%
\bibitem [{\citenamefont {Singh}\ \emph {et~al.}(2013)\citenamefont {Singh},
  \citenamefont {Bhuyan}, \citenamefont {Panda},\ and\ \citenamefont
  {Patra}}]{Singh_2013}%
  \BibitemOpen
  \bibfield  {author} {\bibinfo {author} {\bibfnamefont {S.~K.}\ \bibnamefont
  {Singh}}, \bibinfo {author} {\bibfnamefont {M.}~\bibnamefont {Bhuyan}},
  \bibinfo {author} {\bibfnamefont {P.~K.}\ \bibnamefont {Panda}}, \ and\
  \bibinfo {author} {\bibfnamefont {S.~K.}\ \bibnamefont {Patra}},\ }\href
  {\doibase 10.1088/0954-3899/40/8/085104} {\bibfield  {journal} {\bibinfo
  {journal} {Journal of Physics G: Nuclear and Particle Physics}\ }\textbf
  {\bibinfo {volume} {40}},\ \bibinfo {pages} {085104} (\bibinfo {year}
  {2013})},\ \Eprint {http://arxiv.org/abs/1211.5461} {arXiv:1211.5461
  [nucl-th]} \BibitemShut {NoStop}%
\bibitem [{\citenamefont {Adhikari}\ \emph {et~al.}(2021)\citenamefont
  {Adhikari}, \citenamefont {Albataineh}, \citenamefont {Androic} \emph
  {et~al.}}]{Adhikari_2021}%
  \BibitemOpen
  \bibfield  {author} {\bibinfo {author} {\bibfnamefont {D.}~\bibnamefont
  {Adhikari}}, \bibinfo {author} {\bibfnamefont {H.}~\bibnamefont
  {Albataineh}}, \bibinfo {author} {\bibfnamefont {D.}~\bibnamefont {Androic}},
   \emph {et~al.} (\bibinfo {collaboration} {PREX Collaboration}),\ }\href
  {\doibase 10.1103/PhysRevLett.126.172502} {\bibfield  {journal} {\bibinfo
  {journal} {Phys. Rev. Lett.}\ }\textbf {\bibinfo {volume} {126}},\ \bibinfo
  {pages} {172502} (\bibinfo {year} {2021})}\BibitemShut {NoStop}%
\bibitem [{\citenamefont {Reed}\ \emph {et~al.}(2021)\citenamefont {Reed},
  \citenamefont {Fattoyev}, \citenamefont {Horowitz},\ and\ \citenamefont
  {Piekarewicz}}]{Reed_2021}%
  \BibitemOpen
  \bibfield  {author} {\bibinfo {author} {\bibfnamefont {B.~T.}\ \bibnamefont
  {Reed}}, \bibinfo {author} {\bibfnamefont {F.~J.}\ \bibnamefont {Fattoyev}},
  \bibinfo {author} {\bibfnamefont {C.~J.}\ \bibnamefont {Horowitz}}, \ and\
  \bibinfo {author} {\bibfnamefont {J.}~\bibnamefont {Piekarewicz}},\ }\href
  {\doibase 10.1103/PhysRevLett.126.172503} {\bibfield  {journal} {\bibinfo
  {journal} {Phys. Rev. Lett.}\ }\textbf {\bibinfo {volume} {126}},\ \bibinfo
  {pages} {172503} (\bibinfo {year} {2021})}\BibitemShut {NoStop}%
\bibitem [{\citenamefont {Miller}\ \emph {et~al.}(2021)\citenamefont {Miller},
  \citenamefont {Lamb}, \citenamefont {Dittmann} \emph {et~al.}}]{Miller_2021}%
  \BibitemOpen
  \bibfield  {author} {\bibinfo {author} {\bibfnamefont {M.~C.}\ \bibnamefont
  {Miller}}, \bibinfo {author} {\bibfnamefont {F.~K.}\ \bibnamefont {Lamb}},
  \bibinfo {author} {\bibfnamefont {A.~J.}\ \bibnamefont {Dittmann}},  \emph
  {et~al.},\ }\href@noop {} {} (\bibinfo {year} {2021}),\ \Eprint
  {http://arxiv.org/abs/2105.06979} {arXiv:2105.06979 [astro-ph.HE]}
  \BibitemShut {NoStop}%
\bibitem [{\citenamefont {Pattnaik}\ \emph {et~al.}(2021)\citenamefont
  {Pattnaik}, \citenamefont {Panda}, \citenamefont {Bhuyan},\ and\
  \citenamefont {Patra}}]{Pattnaik_2021}%
  \BibitemOpen
  \bibfield  {author} {\bibinfo {author} {\bibfnamefont {J.~A.}\ \bibnamefont
  {Pattnaik}}, \bibinfo {author} {\bibfnamefont {R.~N.}\ \bibnamefont {Panda}},
  \bibinfo {author} {\bibfnamefont {M.}~\bibnamefont {Bhuyan}}, \ and\ \bibinfo
  {author} {\bibfnamefont {S.~K.}\ \bibnamefont {Patra}},\ }\href@noop {} {}
  (\bibinfo {year} {2021}),\ \Eprint {http://arxiv.org/abs/2105.14479}
  {arXiv:2105.14479 [nucl-th]} \BibitemShut {NoStop}%
\bibitem [{\citenamefont {Zhang}\ and\ \citenamefont
  {Chen}(2013)}]{ZHANG_2013}%
  \BibitemOpen
  \bibfield  {author} {\bibinfo {author} {\bibfnamefont {Z.}~\bibnamefont
  {Zhang}}\ and\ \bibinfo {author} {\bibfnamefont {L.-W.}\ \bibnamefont
  {Chen}},\ }\href {\doibase https://doi.org/10.1016/j.physletb.2013.08.002}
  {\bibfield  {journal} {\bibinfo  {journal} {Physics Letters B}\ }\textbf
  {\bibinfo {volume} {726}},\ \bibinfo {pages} {234} (\bibinfo {year}
  {2013})}\BibitemShut {NoStop}%
\bibitem [{\citenamefont {Horowitz}\ and\ \citenamefont
  {Piekarewicz}(2001)}]{Horowitz_2001}%
  \BibitemOpen
  \bibfield  {author} {\bibinfo {author} {\bibfnamefont {C.~J.}\ \bibnamefont
  {Horowitz}}\ and\ \bibinfo {author} {\bibfnamefont {J.}~\bibnamefont
  {Piekarewicz}},\ }\href {\doibase 10.1103/PhysRevLett.86.5647} {\bibfield
  {journal} {\bibinfo  {journal} {Phys. Rev. Lett.}\ }\textbf {\bibinfo
  {volume} {86}},\ \bibinfo {pages} {5647} (\bibinfo {year}
  {2001})}\BibitemShut {NoStop}%
\bibitem [{\citenamefont {Ducoin}\ \emph {et~al.}(2010)\citenamefont {Ducoin},
  \citenamefont {Margueron},\ and\ \citenamefont
  {Provid{\^{e}}ncia}}]{Ducoin_2010}%
  \BibitemOpen
  \bibfield  {author} {\bibinfo {author} {\bibfnamefont {C.}~\bibnamefont
  {Ducoin}}, \bibinfo {author} {\bibfnamefont {J.}~\bibnamefont {Margueron}}, \
  and\ \bibinfo {author} {\bibfnamefont {C.}~\bibnamefont
  {Provid{\^{e}}ncia}},\ }\href {\doibase 10.1209/0295-5075/91/32001}
  {\bibfield  {journal} {\bibinfo  {journal} {{EPL} (Europhysics Letters)}\
  }\textbf {\bibinfo {volume} {91}},\ \bibinfo {pages} {32001} (\bibinfo {year}
  {2010})}\BibitemShut {NoStop}%
\bibitem [{\citenamefont {Ducoin}\ \emph {et~al.}(2011)\citenamefont {Ducoin},
  \citenamefont {Margueron}, \citenamefont {Provid\^encia},\ and\ \citenamefont
  {Vida\~na}}]{Ducoin_2011}%
  \BibitemOpen
  \bibfield  {author} {\bibinfo {author} {\bibfnamefont {C.}~\bibnamefont
  {Ducoin}}, \bibinfo {author} {\bibfnamefont {J.}~\bibnamefont {Margueron}},
  \bibinfo {author} {\bibfnamefont {C.~m.~c.}\ \bibnamefont {Provid\^encia}}, \
  and\ \bibinfo {author} {\bibfnamefont {I.}~\bibnamefont {Vida\~na}},\ }\href
  {\doibase 10.1103/PhysRevC.83.045810} {\bibfield  {journal} {\bibinfo
  {journal} {Phys. Rev. C}\ }\textbf {\bibinfo {volume} {83}},\ \bibinfo
  {pages} {045810} (\bibinfo {year} {2011})}\BibitemShut {NoStop}%
\bibitem [{\citenamefont {Hebeler}\ \emph {et~al.}(2013)\citenamefont
  {Hebeler}, \citenamefont {Lattimer}, \citenamefont {Pethick},\ and\
  \citenamefont {Schwenk}}]{Hebeler_2013}%
  \BibitemOpen
  \bibfield  {author} {\bibinfo {author} {\bibfnamefont {K.}~\bibnamefont
  {Hebeler}}, \bibinfo {author} {\bibfnamefont {J.~M.}\ \bibnamefont
  {Lattimer}}, \bibinfo {author} {\bibfnamefont {C.~J.}\ \bibnamefont
  {Pethick}}, \ and\ \bibinfo {author} {\bibfnamefont {A.}~\bibnamefont
  {Schwenk}},\ }\href {\doibase 10.1088/0004-637x/773/1/11} {\bibfield
  {journal} {\bibinfo  {journal} {The Astrophysical Journal}\ }\textbf
  {\bibinfo {volume} {773}},\ \bibinfo {pages} {11} (\bibinfo {year}
  {2013})}\BibitemShut {NoStop}%
\bibitem [{\citenamefont {Horowitz}\ \emph {et~al.}(2014)\citenamefont
  {Horowitz}, \citenamefont {Brown}, \citenamefont {Kim} \emph
  {et~al.}}]{Horowitz_2014}%
  \BibitemOpen
  \bibfield  {author} {\bibinfo {author} {\bibfnamefont {C.~J.}\ \bibnamefont
  {Horowitz}}, \bibinfo {author} {\bibfnamefont {E.~F.}\ \bibnamefont {Brown}},
  \bibinfo {author} {\bibfnamefont {Y.}~\bibnamefont {Kim}},  \emph {et~al.},\
  }\href {\doibase 10.1088/0954-3899/41/9/093001} {\bibfield  {journal}
  {\bibinfo  {journal} {Journal of Physics G: Nuclear and Particle Physics}\
  }\textbf {\bibinfo {volume} {41}},\ \bibinfo {pages} {093001} (\bibinfo
  {year} {2014})}\BibitemShut {NoStop}%
\bibitem [{\citenamefont {Drischler}\ \emph {et~al.}(2020)\citenamefont
  {Drischler}, \citenamefont {Furnstahl}, \citenamefont {Melendez},\ and\
  \citenamefont {Phillips}}]{Drischler_2020}%
  \BibitemOpen
  \bibfield  {author} {\bibinfo {author} {\bibfnamefont {C.}~\bibnamefont
  {Drischler}}, \bibinfo {author} {\bibfnamefont {R.~J.}\ \bibnamefont
  {Furnstahl}}, \bibinfo {author} {\bibfnamefont {J.~A.}\ \bibnamefont
  {Melendez}}, \ and\ \bibinfo {author} {\bibfnamefont {D.~R.}\ \bibnamefont
  {Phillips}},\ }\href {\doibase 10.1103/PhysRevLett.125.202702} {\bibfield
  {journal} {\bibinfo  {journal} {Phys. Rev. Lett.}\ }\textbf {\bibinfo
  {volume} {125}},\ \bibinfo {pages} {202702} (\bibinfo {year}
  {2020})}\BibitemShut {NoStop}%
\bibitem [{\citenamefont {Hagen}\ \emph {et~al.}(2016)\citenamefont {Hagen},
  \citenamefont {Ekstr{\"o}m}, \citenamefont {Forss{\'e}n} \emph
  {et~al.}}]{Hagen_2016}%
  \BibitemOpen
  \bibfield  {author} {\bibinfo {author} {\bibfnamefont {G.}~\bibnamefont
  {Hagen}}, \bibinfo {author} {\bibfnamefont {A.}~\bibnamefont {Ekstr{\"o}m}},
  \bibinfo {author} {\bibfnamefont {C.}~\bibnamefont {Forss{\'e}n}},  \emph
  {et~al.},\ }\href {\doibase 10.1038/nphys3529} {\bibfield  {journal}
  {\bibinfo  {journal} {Nature Physics}\ }\textbf {\bibinfo {volume} {12}},\
  \bibinfo {pages} {186} (\bibinfo {year} {2016})}\BibitemShut {NoStop}%
\bibitem [{\citenamefont {Chen}\ \emph {et~al.}(2010)\citenamefont {Chen},
  \citenamefont {Ko}, \citenamefont {Li},\ and\ \citenamefont
  {Xu}}]{Chen_2010}%
  \BibitemOpen
  \bibfield  {author} {\bibinfo {author} {\bibfnamefont {L.-W.}\ \bibnamefont
  {Chen}}, \bibinfo {author} {\bibfnamefont {C.~M.}\ \bibnamefont {Ko}},
  \bibinfo {author} {\bibfnamefont {B.-A.}\ \bibnamefont {Li}}, \ and\ \bibinfo
  {author} {\bibfnamefont {J.}~\bibnamefont {Xu}},\ }\href {\doibase
  10.1103/PhysRevC.82.024321} {\bibfield  {journal} {\bibinfo  {journal} {Phys.
  Rev. C}\ }\textbf {\bibinfo {volume} {82}},\ \bibinfo {pages} {024321}
  (\bibinfo {year} {2010})}\BibitemShut {NoStop}%
\bibitem [{\citenamefont {Gandolfi}\ \emph {et~al.}(2014)\citenamefont
  {Gandolfi}, \citenamefont {Carlson}, \citenamefont {Reddy}, \citenamefont
  {Steiner},\ and\ \citenamefont {Wiringa}}]{Gandolfi_2014}%
  \BibitemOpen
  \bibfield  {author} {\bibinfo {author} {\bibfnamefont {S.}~\bibnamefont
  {Gandolfi}}, \bibinfo {author} {\bibfnamefont {J.}~\bibnamefont {Carlson}},
  \bibinfo {author} {\bibfnamefont {S.}~\bibnamefont {Reddy}}, \bibinfo
  {author} {\bibfnamefont {A.~W.}\ \bibnamefont {Steiner}}, \ and\ \bibinfo
  {author} {\bibfnamefont {R.~B.}\ \bibnamefont {Wiringa}},\ }\href {\doibase
  10.1140/epja/i2014-14010-5} {\bibfield  {journal} {\bibinfo  {journal} {The
  European Physical Journal A}\ }\textbf {\bibinfo {volume} {50}},\ \bibinfo
  {pages} {10} (\bibinfo {year} {2014})}\BibitemShut {NoStop}%
\bibitem [{\citenamefont {Roca-Maza}\ \emph {et~al.}(2015)\citenamefont
  {Roca-Maza}, \citenamefont {Vi\~nas}, \citenamefont {Centelles},
  \citenamefont {Agrawal}, \citenamefont {Col\`o}, \citenamefont {Paar},
  \citenamefont {Piekarewicz},\ and\ \citenamefont {Vretenar}}]{Roca_2015}%
  \BibitemOpen
  \bibfield  {author} {\bibinfo {author} {\bibfnamefont {X.}~\bibnamefont
  {Roca-Maza}}, \bibinfo {author} {\bibfnamefont {X.}~\bibnamefont {Vi\~nas}},
  \bibinfo {author} {\bibfnamefont {M.}~\bibnamefont {Centelles}}, \bibinfo
  {author} {\bibfnamefont {B.~K.}\ \bibnamefont {Agrawal}}, \bibinfo {author}
  {\bibfnamefont {G.}~\bibnamefont {Col\`o}}, \bibinfo {author} {\bibfnamefont
  {N.}~\bibnamefont {Paar}}, \bibinfo {author} {\bibfnamefont {J.}~\bibnamefont
  {Piekarewicz}}, \ and\ \bibinfo {author} {\bibfnamefont {D.}~\bibnamefont
  {Vretenar}},\ }\href {\doibase 10.1103/PhysRevC.92.064304} {\bibfield
  {journal} {\bibinfo  {journal} {Phys. Rev. C}\ }\textbf {\bibinfo {volume}
  {92}},\ \bibinfo {pages} {064304} (\bibinfo {year} {2015})}\BibitemShut
  {NoStop}%
\bibitem [{\citenamefont {Li}\ \emph {et~al.}(2019)\citenamefont {Li},
  \citenamefont {Krastev}, \citenamefont {Wen},\ and\ \citenamefont
  {Zhang}}]{BaoLi_2019}%
  \BibitemOpen
  \bibfield  {author} {\bibinfo {author} {\bibfnamefont {B.-A.}\ \bibnamefont
  {Li}}, \bibinfo {author} {\bibfnamefont {P.~G.}\ \bibnamefont {Krastev}},
  \bibinfo {author} {\bibfnamefont {D.-H.}\ \bibnamefont {Wen}}, \ and\
  \bibinfo {author} {\bibfnamefont {N.-B.}\ \bibnamefont {Zhang}},\ }\href
  {\doibase 10.1140/epja/i2019-12780-8} {\bibfield  {journal} {\bibinfo
  {journal} {EPJ A}\ }\textbf {\bibinfo {volume} {55}} (\bibinfo {year}
  {2019}),\ 10.1140/epja/i2019-12780-8}\BibitemShut {NoStop}%
\bibitem [{\citenamefont {Li}\ and\ \citenamefont {Han}(2013)}]{BaoLi_2013}%
  \BibitemOpen
  \bibfield  {author} {\bibinfo {author} {\bibfnamefont {B.-A.}\ \bibnamefont
  {Li}}\ and\ \bibinfo {author} {\bibfnamefont {X.}~\bibnamefont {Han}},\
  }\href {\doibase https://doi.org/10.1016/j.physletb.2013.10.006} {\bibfield
  {journal} {\bibinfo  {journal} {Physics Letters B}\ }\textbf {\bibinfo
  {volume} {727}},\ \bibinfo {pages} {276 } (\bibinfo {year}
  {2013})}\BibitemShut {NoStop}%
\bibitem [{\citenamefont {Matsui}(1981)}]{Matsui_1981}%
  \BibitemOpen
  \bibfield  {author} {\bibinfo {author} {\bibfnamefont {T.}~\bibnamefont
  {Matsui}},\ }\href {\doibase https://doi.org/10.1016/0375-9474(81)90103-2}
  {\bibfield  {journal} {\bibinfo  {journal} {Nucl. Phys. A}\ }\textbf
  {\bibinfo {volume} {370}},\ \bibinfo {pages} {365 } (\bibinfo {year}
  {1981})}\BibitemShut {NoStop}%
\bibitem [{\citenamefont {Kubis}\ and\ \citenamefont
  {Kutschera}(1997)}]{Kubis_1997}%
  \BibitemOpen
  \bibfield  {author} {\bibinfo {author} {\bibfnamefont {S.}~\bibnamefont
  {Kubis}}\ and\ \bibinfo {author} {\bibfnamefont {M.}~\bibnamefont
  {Kutschera}},\ }\href {\doibase 10.1016/s0370-2693(97)00306-7} {\bibfield
  {journal} {\bibinfo  {journal} {Phys. Lett. B}\ }\textbf {\bibinfo {volume}
  {399}},\ \bibinfo {pages} {191–195} (\bibinfo {year} {1997})}\BibitemShut
  {NoStop}%
\bibitem [{\citenamefont {Del~Estal}\ \emph {et~al.}(2001)\citenamefont
  {Del~Estal}, \citenamefont {Centelles}, \citenamefont {Vi\~nas},\ and\
  \citenamefont {Patra}}]{MCentelles_2001}%
  \BibitemOpen
  \bibfield  {author} {\bibinfo {author} {\bibfnamefont {M.}~\bibnamefont
  {Del~Estal}}, \bibinfo {author} {\bibfnamefont {M.}~\bibnamefont
  {Centelles}}, \bibinfo {author} {\bibfnamefont {X.}~\bibnamefont {Vi\~nas}},
  \ and\ \bibinfo {author} {\bibfnamefont {S.~K.}\ \bibnamefont {Patra}},\
  }\href {\doibase 10.1103/PhysRevC.63.024314} {\bibfield  {journal} {\bibinfo
  {journal} {Phys. Rev. C}\ }\textbf {\bibinfo {volume} {63}},\ \bibinfo
  {pages} {024314} (\bibinfo {year} {2001})}\BibitemShut {NoStop}%
\bibitem [{\citenamefont {Chen}\ and\ \citenamefont
  {Piekarewicz}(2014)}]{Chen_2014}%
  \BibitemOpen
  \bibfield  {author} {\bibinfo {author} {\bibfnamefont {W.-C.}\ \bibnamefont
  {Chen}}\ and\ \bibinfo {author} {\bibfnamefont {J.}~\bibnamefont
  {Piekarewicz}},\ }\href {\doibase 10.1103/PhysRevC.90.044305} {\bibfield
  {journal} {\bibinfo  {journal} {Phys. Rev. C}\ }\textbf {\bibinfo {volume}
  {90}},\ \bibinfo {pages} {044305} (\bibinfo {year} {2014})}\BibitemShut
  {NoStop}%
\bibitem [{\citenamefont {Fortin}\ \emph {et~al.}(2016)\citenamefont {Fortin},
  \citenamefont {Provid\^encia}, \citenamefont {Raduta}, \citenamefont
  {Gulminelli}, \citenamefont {Zdunik}, \citenamefont {Haensel},\ and\
  \citenamefont {Bejger}}]{Fortin_2016}%
  \BibitemOpen
  \bibfield  {author} {\bibinfo {author} {\bibfnamefont {M.}~\bibnamefont
  {Fortin}}, \bibinfo {author} {\bibfnamefont {C.}~\bibnamefont
  {Provid\^encia}}, \bibinfo {author} {\bibfnamefont {A.~R.}\ \bibnamefont
  {Raduta}}, \bibinfo {author} {\bibfnamefont {F.}~\bibnamefont {Gulminelli}},
  \bibinfo {author} {\bibfnamefont {J.~L.}\ \bibnamefont {Zdunik}}, \bibinfo
  {author} {\bibfnamefont {P.}~\bibnamefont {Haensel}}, \ and\ \bibinfo
  {author} {\bibfnamefont {M.}~\bibnamefont {Bejger}},\ }\href {\doibase
  10.1103/PhysRevC.94.035804} {\bibfield  {journal} {\bibinfo  {journal} {Phys.
  Rev. C}\ }\textbf {\bibinfo {volume} {94}},\ \bibinfo {pages} {035804}
  (\bibinfo {year} {2016})}\BibitemShut {NoStop}%
\bibitem [{\citenamefont {Biswal}\ \emph {et~al.}(2020)\citenamefont {Biswal},
  \citenamefont {Das}, \citenamefont {Kumar}, \citenamefont {Kumar},\ and\
  \citenamefont {Patra}}]{Biswal_2020}%
  \BibitemOpen
  \bibfield  {author} {\bibinfo {author} {\bibfnamefont {S.~K.}\ \bibnamefont
  {Biswal}}, \bibinfo {author} {\bibfnamefont {H.~C.}\ \bibnamefont {Das}},
  \bibinfo {author} {\bibfnamefont {A.}~\bibnamefont {Kumar}}, \bibinfo
  {author} {\bibfnamefont {B.}~\bibnamefont {Kumar}}, \ and\ \bibinfo {author}
  {\bibfnamefont {S.~K.}\ \bibnamefont {Patra}},\ }\href@noop {} {\  (\bibinfo
  {year} {2020})},\ \Eprint {http://arxiv.org/abs/2012.13673} {arXiv:2012.13673
  [astro-ph.HE]} \BibitemShut {NoStop}%
\bibitem [{\citenamefont {Roca-Maza}\ \emph {et~al.}(2011)\citenamefont
  {Roca-Maza}, \citenamefont {Centelles}, \citenamefont {Vi\~nas},\ and\
  \citenamefont {Warda}}]{Roca_2011}%
  \BibitemOpen
  \bibfield  {author} {\bibinfo {author} {\bibfnamefont {X.}~\bibnamefont
  {Roca-Maza}}, \bibinfo {author} {\bibfnamefont {M.}~\bibnamefont
  {Centelles}}, \bibinfo {author} {\bibfnamefont {X.}~\bibnamefont {Vi\~nas}},
  \ and\ \bibinfo {author} {\bibfnamefont {M.}~\bibnamefont {Warda}},\ }\href
  {\doibase 10.1103/PhysRevLett.106.252501} {\bibfield  {journal} {\bibinfo
  {journal} {Phys. Rev. Lett.}\ }\textbf {\bibinfo {volume} {106}},\ \bibinfo
  {pages} {252501} (\bibinfo {year} {2011})}\BibitemShut {NoStop}%
\bibitem [{\citenamefont {Das}\ \emph {et~al.}(2020)\citenamefont {Das},
  \citenamefont {Kumar}, \citenamefont {Kumar} \emph {et~al.}}]{Das_2020}%
  \BibitemOpen
  \bibfield  {author} {\bibinfo {author} {\bibfnamefont {H.~C.}\ \bibnamefont
  {Das}}, \bibinfo {author} {\bibfnamefont {A.}~\bibnamefont {Kumar}}, \bibinfo
  {author} {\bibfnamefont {B.}~\bibnamefont {Kumar}},  \emph {et~al.},\ }\href
  {\doibase 10.1093/mnras/staa1435} {\bibfield  {journal} {\bibinfo  {journal}
  {MNRAS}\ }\textbf {\bibinfo {volume} {495}},\ \bibinfo {pages} {4893}
  (\bibinfo {year} {2020})}\BibitemShut {NoStop}%
\bibitem [{\citenamefont {Das}\ \emph {et~al.}(2021{\natexlab{a}})\citenamefont
  {Das}, \citenamefont {Kumar}, \citenamefont {Kumar}, \citenamefont {Biswal},\
  and\ \citenamefont {Patra}}]{Das_2021}%
  \BibitemOpen
  \bibfield  {author} {\bibinfo {author} {\bibfnamefont {H.~C.}\ \bibnamefont
  {Das}}, \bibinfo {author} {\bibfnamefont {A.}~\bibnamefont {Kumar}}, \bibinfo
  {author} {\bibfnamefont {B.}~\bibnamefont {Kumar}}, \bibinfo {author}
  {\bibfnamefont {S.~K.}\ \bibnamefont {Biswal}}, \ and\ \bibinfo {author}
  {\bibfnamefont {S.~K.}\ \bibnamefont {Patra}},\ }\href {\doibase
  10.1088/1475-7516/2021/01/007} {\bibfield  {journal} {\bibinfo  {journal}
  {JCAP}\ }\textbf {\bibinfo {volume} {2021}},\ \bibinfo {pages} {007}
  (\bibinfo {year} {2021}{\natexlab{a}})}\BibitemShut {NoStop}%
\bibitem [{\citenamefont {Das}\ \emph {et~al.}(2021{\natexlab{b}})\citenamefont
  {Das}, \citenamefont {Kumar},\ and\ \citenamefont {Patra}}]{das2021effects}%
  \BibitemOpen
  \bibfield  {author} {\bibinfo {author} {\bibfnamefont {H.~C.}\ \bibnamefont
  {Das}}, \bibinfo {author} {\bibfnamefont {A.}~\bibnamefont {Kumar}}, \ and\
  \bibinfo {author} {\bibfnamefont {S.~K.}\ \bibnamefont {Patra}},\ }\href@noop
  {} {} (\bibinfo {year} {2021}{\natexlab{b}}),\ \Eprint
  {http://arxiv.org/abs/2104.01815} {arXiv:2104.01815 [astro-ph.HE]}
  \BibitemShut {NoStop}%
\bibitem [{\citenamefont {{Tolman}}(1939)}]{TOV1}%
  \BibitemOpen
  \bibfield  {author} {\bibinfo {author} {\bibfnamefont {R.~C.}\ \bibnamefont
  {{Tolman}}},\ }\href {\doibase 10.1103/PhysRev.55.364} {\bibfield  {journal}
  {\bibinfo  {journal} {Phys. Rev.}\ }\textbf {\bibinfo {volume} {55}},\
  \bibinfo {pages} {364} (\bibinfo {year} {1939})}\BibitemShut {NoStop}%
\bibitem [{\citenamefont {Oppenheimer}\ and\ \citenamefont
  {Volkoff}(1939)}]{TOV2}%
  \BibitemOpen
  \bibfield  {author} {\bibinfo {author} {\bibfnamefont {J.~R.}\ \bibnamefont
  {Oppenheimer}}\ and\ \bibinfo {author} {\bibfnamefont {G.~M.}\ \bibnamefont
  {Volkoff}},\ }\href {\doibase 10.1103/PhysRev.55.374} {\bibfield  {journal}
  {\bibinfo  {journal} {Phys. Rev.}\ }\textbf {\bibinfo {volume} {55}},\
  \bibinfo {pages} {374} (\bibinfo {year} {1939})}\BibitemShut {NoStop}%
\bibitem [{\citenamefont {Demorest}\ \emph {et~al.}(2010)\citenamefont
  {Demorest}, \citenamefont {Pennucci}, \citenamefont {Ransom}, \citenamefont
  {Roberts},\ and\ \citenamefont {Hessels}}]{Demorest_2010}%
  \BibitemOpen
  \bibfield  {author} {\bibinfo {author} {\bibfnamefont {P.~B.}\ \bibnamefont
  {Demorest}}, \bibinfo {author} {\bibfnamefont {T.}~\bibnamefont {Pennucci}},
  \bibinfo {author} {\bibfnamefont {S.~M.}\ \bibnamefont {Ransom}}, \bibinfo
  {author} {\bibfnamefont {M.~S.~E.}\ \bibnamefont {Roberts}}, \ and\ \bibinfo
  {author} {\bibfnamefont {J.~W.~T.}\ \bibnamefont {Hessels}},\ }\href
  {\doibase 10.1038/nature09466} {\bibfield  {journal} {\bibinfo  {journal}
  {Nature}\ }\textbf {\bibinfo {volume} {467}},\ \bibinfo {pages} {1081–1083}
  (\bibinfo {year} {2010})}\BibitemShut {NoStop}%
\bibitem [{\citenamefont {Antoniadis}\ \emph {et~al.}(2013)\citenamefont
  {Antoniadis}, \citenamefont {Freire} \emph {et~al.}}]{Antoniadis_2013}%
  \BibitemOpen
  \bibfield  {author} {\bibinfo {author} {\bibfnamefont {J.}~\bibnamefont
  {Antoniadis}}, \bibinfo {author} {\bibfnamefont {P.~C.~C.}\ \bibnamefont
  {Freire}},  \emph {et~al.},\ }\href {\doibase 10.1126/science.1233232}
  {\bibfield  {journal} {\bibinfo  {journal} {Science}\ }\textbf {\bibinfo
  {volume} {340}} (\bibinfo {year} {2013}),\
  10.1126/science.1233232}\BibitemShut {NoStop}%
\bibitem [{\citenamefont {Cromartie}\ \emph {et~al.}(2019)\citenamefont
  {Cromartie}, \citenamefont {Fonseca}, \citenamefont {Ransom}, \citenamefont
  {Demorest} \emph {et~al.}}]{Cromartie_2019}%
  \BibitemOpen
  \bibfield  {author} {\bibinfo {author} {\bibfnamefont {H.~T.}\ \bibnamefont
  {Cromartie}}, \bibinfo {author} {\bibfnamefont {E.}~\bibnamefont {Fonseca}},
  \bibinfo {author} {\bibfnamefont {S.~M.}\ \bibnamefont {Ransom}}, \bibinfo
  {author} {\bibfnamefont {P.~B.}\ \bibnamefont {Demorest}},  \emph {et~al.},\
  }\href {\doibase 10.1038/s41550-019-0880-2} {\bibfield  {journal} {\bibinfo
  {journal} {Nature Astronomy}\ }\textbf {\bibinfo {volume} {4}},\ \bibinfo
  {pages} {72–76} (\bibinfo {year} {2019})}\BibitemShut {NoStop}%
\bibitem [{\citenamefont {Abbott}\ \emph {et~al.}(2020)\citenamefont {Abbott},
  \citenamefont {Abbott}, \citenamefont {Abraham} \emph
  {et~al.}}]{RAbbott_2020}%
  \BibitemOpen
  \bibfield  {author} {\bibinfo {author} {\bibfnamefont {R.}~\bibnamefont
  {Abbott}}, \bibinfo {author} {\bibfnamefont {T.~D.}\ \bibnamefont {Abbott}},
  \bibinfo {author} {\bibfnamefont {S.}~\bibnamefont {Abraham}},  \emph
  {et~al.},\ }\href {\doibase 10.3847/2041-8213/ab960f} {\bibfield  {journal}
  {\bibinfo  {journal} {The Astrophysical Journal}\ }\textbf {\bibinfo {volume}
  {896}},\ \bibinfo {pages} {L44} (\bibinfo {year} {2020})},\ \Eprint
  {http://arxiv.org/abs/2006.12611} {arXiv:2006.12611 [astro-ph.HE]}
  \BibitemShut {NoStop}%
\bibitem [{\citenamefont {Das}\ \emph {et~al.}(2021{\natexlab{c}})\citenamefont
  {Das}, \citenamefont {Kumar},\ and\ \citenamefont {Patra}}]{DasPRD_2021}%
  \BibitemOpen
  \bibfield  {author} {\bibinfo {author} {\bibfnamefont {H.~C.}\ \bibnamefont
  {Das}}, \bibinfo {author} {\bibfnamefont {A.}~\bibnamefont {Kumar}}, \ and\
  \bibinfo {author} {\bibfnamefont {S.~K.}\ \bibnamefont {Patra}},\ }\href@noop
  {} {} (\bibinfo {year} {2021}{\natexlab{c}}),\ \Eprint
  {http://arxiv.org/abs/2109.01853} {arXiv:2109.01853 [astro-ph.HE]}
  \BibitemShut {NoStop}%
\bibitem [{\citenamefont {Miller}\ \emph {et~al.}(2019)\citenamefont {Miller},
  \citenamefont {Lamb}, \citenamefont {Dittmann} \emph {et~al.}}]{Miller_2019}%
  \BibitemOpen
  \bibfield  {author} {\bibinfo {author} {\bibfnamefont {M.~C.}\ \bibnamefont
  {Miller}}, \bibinfo {author} {\bibfnamefont {F.~K.}\ \bibnamefont {Lamb}},
  \bibinfo {author} {\bibfnamefont {A.~J.}\ \bibnamefont {Dittmann}},  \emph
  {et~al.},\ }\href {\doibase 10.3847/2041-8213/ab50c5} {\bibfield  {journal}
  {\bibinfo  {journal} {APJ}\ }\textbf {\bibinfo {volume} {887}},\ \bibinfo
  {pages} {L24} (\bibinfo {year} {2019})}\BibitemShut {NoStop}%
\bibitem [{\citenamefont {Riley}\ \emph {et~al.}(2019)\citenamefont {Riley},
  \citenamefont {Watts}, \citenamefont {Bogdanov} \emph {et~al.}}]{Riley_2019}%
  \BibitemOpen
  \bibfield  {author} {\bibinfo {author} {\bibfnamefont {T.~E.}\ \bibnamefont
  {Riley}}, \bibinfo {author} {\bibfnamefont {A.~L.}\ \bibnamefont {Watts}},
  \bibinfo {author} {\bibfnamefont {S.}~\bibnamefont {Bogdanov}},  \emph
  {et~al.},\ }\href {\doibase 10.3847/2041-8213/ab481c} {\bibfield  {journal}
  {\bibinfo  {journal} {APJ}\ }\textbf {\bibinfo {volume} {887}},\ \bibinfo
  {pages} {L21} (\bibinfo {year} {2019})}\BibitemShut {NoStop}%
\end{thebibliography}%
\bibliographystyle{apsrev4-1}
\end{document}